\documentclass[osajnl,preprint,showpacs,superscriptaddress,12pt]{revtex4-1} 
\usepackage{amsmath,amssymb,graphicx}
\usepackage{xcolor}
\begin{document}

\title{Light Focusing and Two-Dimensional Imaging Through Scattering Media using the Photoacoustic Transmission-Matrix with an Ultrasound Array}

\author{Thomas Chaigne}
\author{J\'er\^ome Gateau}
\author{Ori Katz}
\author{Emmanuel Bossy}\email{Corresponding author: emmanuel.bossy@espci.fr}
\author{Sylvain Gigan}\email{Corresponding author: sylvain.gigan@espci.fr}

\affiliation{Institut Langevin, ESPCI ParisTech, CNRS UMR 7587, INSERM U979, 1 rue Jussieu, 75005 Paris, France}

\begin{abstract}
We implement the photoacoustic transmission-matrix approach on a two-dimensional photoacoustic imaging system, using a 15 MHz linear ultrasound array. Using a black leaf skeleton as a complex absorbing structure, we demonstrate that the photoacoustic transmission-matrix approach allows to reveal structural features that are invisible in conventional photoacoustic images, as well as to selectively control light focusing on absorbing targets, leading to a local enhancement of the photoacoustic signal.

\end{abstract}

\maketitle 

In biological tissue, light scattering limits the penetration depth of most optical imaging techniques to a few hundred micrometers~\cite{ntziachristos2010going}. 
In the last few years, wavefront shaping appeared as a powerful tool to compensate light scattering. By controlling the input wavefront with a spatial light modulator (SLM), Vellekoop and Mosk first demonstrated the focusing of coherent light through highly scattering opaque layers~\cite{vellekoop2007focusing}. 
Popoff et al. then introduced the transmission-matrix approach, allowing imaging and focusing through highly scattering samples, by measuring the complex linear relations between many input and output optical modes~\cite{popoff2010measuring}. Many wavefront-shaping based techniques have been thoroughly investigated since~\cite{mosk2012controlling}, but most of the proposed approach relied on a feedback signal from a  camera placed behind the scattering sample, i.e. at the target plane. Although the requirement for a direct optical access to the target was considered as the main limitation for applying wavefront shaping in practical scenarios, several approaches to overcome this apparent limitation have been investigated. These methods include using implanted guide-stars~\cite{vellekoop2008demixing,hsieh2010imaging}, ultrasonic tagging of light~\cite{xu2011time,si2012fluorescence,judkewitz2013speckle}, and the photoacoustic effect~\cite{kong2011photoacoustic}.\\ 

Photoacoustic imaging has emerged in the past decade as a powerful hybrid modality to image optical-absorption contrast in turbid media~\cite{beard2011biomedical}. Because ultrasound waves are only weakly scattered, images can be obtained at penetration depth up to a few centimeters with a sub-mm resolution. By using the photoacoustic feedback from a single ultrasonic transducer,  Kong et al. performed wavefront shaping to focus light on absorbers located behind a scattering sample~\cite{kong2011photoacoustic}. The use of photoacoustic-guided wavefront shaping has then been extended to allow selective focusing at multiple absorbing targets using the photoacoustic transmission matrix~\cite{chaigne2014controlling}. Photoacoustic-guided wavefront-shaping has also been very recently performed using the large spectral bandwidth of photoacoustic signals~\cite{chaigne2013improving}, and it has been proposed as a mean to get sub-photoacoustic resolution~\cite{2013arXiv1310.5736C}. In addition to providing a feedback signal for wavefront shaping, it appears that coupling photoacoustics and wavefront-shaping could also improve the quality of the photoacoustic images~\cite{caravaca2013high}.\\

In all the above-mentioned works, one-dimensional photoacoustic "images" were acquired using a single spherically focused ultrasonic transducer, thus limiting the applicability of the technique to narrow fields-of-view and requiring mechanical scanning to obtain two-dimensional images. In this Letter, we extend the photoacoustic transmission matrix method~\cite{chaigne2014controlling} to a two-dimensional photoacoustic imaging setup using a linear ultrasound array. Photoacoustic imaging using transducer-arrays is a well established technique which provides a two-dimensional photoacoustic image in a single laser shot. Commercial linear transducer-arrays are widely used in this domain because they enable imaging a whole plane, and benefit from technologies developed for ultrasonic measurements. By combining wavefront shaping and photoacoustic imaging with an ultrasound array, we propose to measure the photoacoustic transmission-matrix over a large field-of-view without scanning. With two-dimensional photoacoustic images, the photoacoustic transmission matrix describes the complex influence of the optical input modes (SLM pixels) on the acoustic output modes (defined as the acoustic cells~\cite{gateau2013improving}, i.e. the intersection between each acoustic resolution cell in the photoacoustic image and the absorbing structures~\cite{chaigne2014controlling}). We study the extended photoacoustic imaging capability provided by this matrix, which measurement involve illuminating the sample with multiple speckle patterns. Light focusing and photoacoustic imaging enhancement is demonstrated through scattering phantoms on several target positions on a complex absorbing structure. 
\newline

\begin{figure}[h!]
\centerline{\includegraphics[width=1\columnwidth]{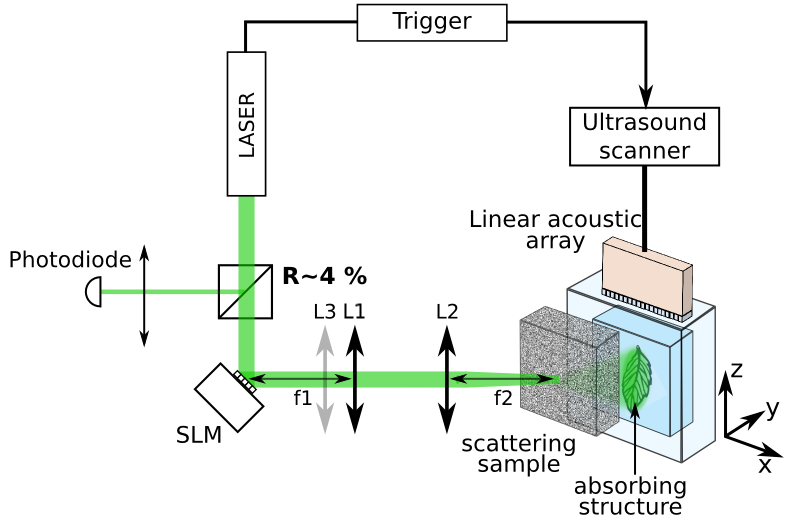}}
\caption{Experimental setup: a 5-ns laser pulse is shaped by an SLM before passing through a scattering sample and illuminating an absorbing structure. Photoacoustic signals are detected with an ultrasonic linear array, driven by an ultrasound scanner.}
\label{montage}
\end{figure}

\begin{figure}[h!]
\centerline{\includegraphics[width=1\columnwidth]{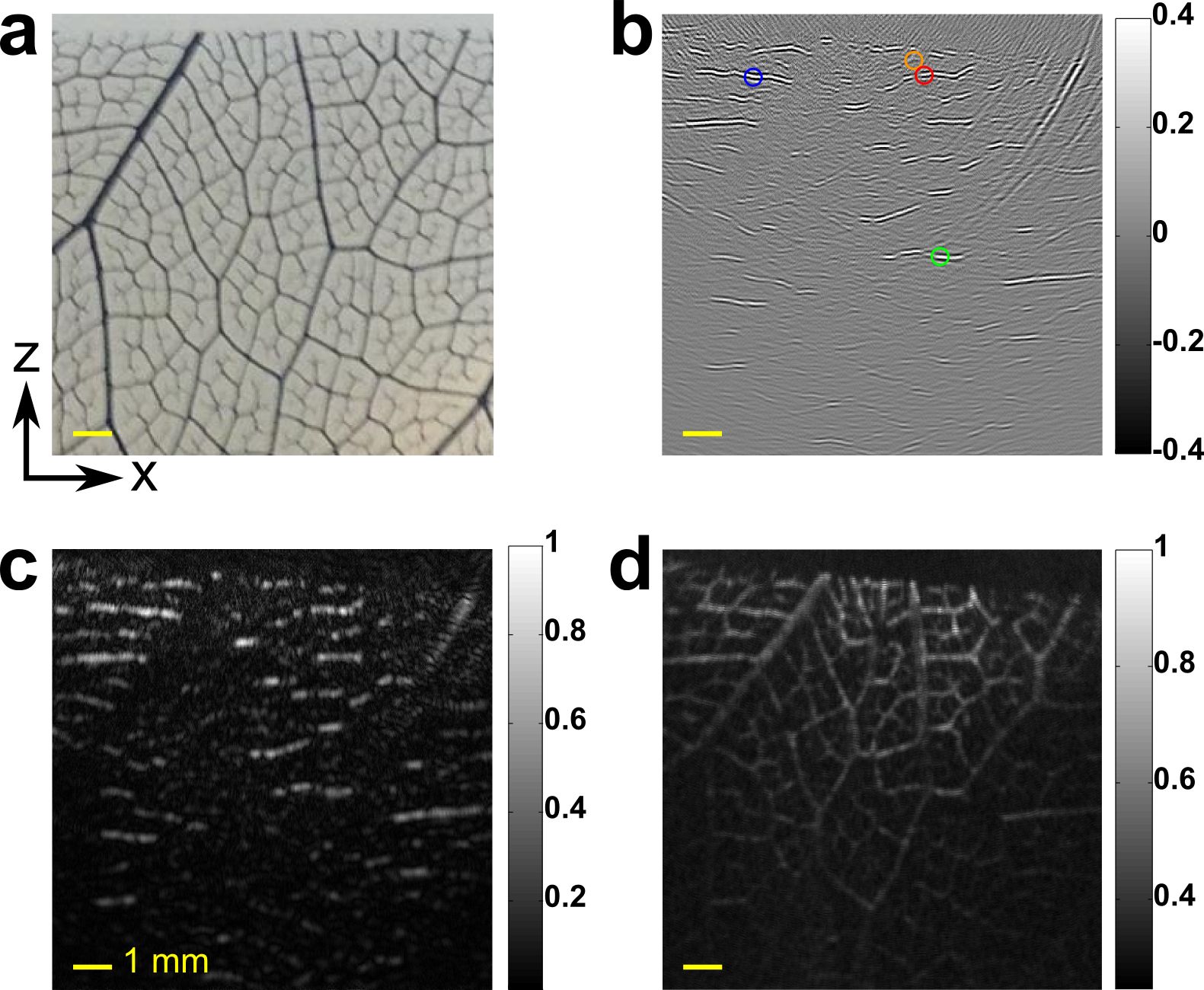}}
\caption{Improved photoacoustic imaging using the transmission matrix: (a) Reference photograph of the absorbing leaf skeleton embedded in an agarose gel block. (b) A standard photoacoustic image equivalent to that obtained with uniform illumination:  this image is obtained by averaging all the photoacoustic images acquired during the transmission matrix measurement, that correspond to the hundreds of speckle illuminations created by the successive SLM phase patterns. This image contains negative values because of the limited bandwidth of the imaging system and artefacts of the backprojection algorithm. The coloured circles indicate the position of the foci presented in the next figures. (c) Envelope image derived from (b), enabling a better visualization analog to that of conventional ultrasound images. (d) Modulation image: the value of each pixel gives the mean modulation depth of the photoacoustic signals (modulus of the transmission matrix), averaged over the input modes.}
\label{leaf}
\end{figure}

The experimental setup is illustrated in Fig.\ref{montage}. A laser pulse (Continuum Surelite, 532 nm wavelength, 5 ns pulse duration, 10 Hz repetition rate) is spatially shaped by an SLM (Multi-DM, Boston Micromachines, 140 segmented mirrors). The spatially-shaped pulse passes through a scattering sample (120 Grit ground glass diffuser, Thorlabs). The SLM plane is conjugated with the scattering surface by a 4-f telescope. This ensures that the optical speckle size on the target remains constant for different SLM phase-patterns~\cite{vellekoop2010exploiting}. An third lens adds a curvature on top of the SLM phase pattern~\cite{chaigne2014controlling}. The scattered light illuminates a phantom containing an absorbing structure, embedded in an agarose gel with negligible optical scattering. In this study, we use a black leaf skeleton as the absorbing target (Fig.\ref{leaf}.a)~\cite{jose2013speed,huang2013improving}. This sample contains branching structures with multiple orientations and various spatial scales, ranging from $50 \mu m$ to $200 \mu m$. This two-dimensional sample is located in the imaging plane and in the vicinity of the elevation focus of a linear ultrasound array (128 elements, 14.4 MHz center frequency and pulse-echo -6 dB bandwidth of 6.8 MHz, 0.1 mm pitch and elevation focus of 8 mm, Vermon, France), connected to a commercial ultrasound scanner (Aixplorer, Supersonic Imagine, France). \\

After each laser shot, the photoacoustically induced ultrasonic waves are measured simultaneously by all the elements of the array. The intensity fluctuations of the laser are monitored with a photodiode, and are used to normalize the measured photoacoustic signals. The acquired photoacoustic signals are averaged over 10 laser-pulses to improve the signal-to-noise ratio (SNR). The averaged signals are then filtered with a digital third-order Butterworth lowpass filter having a cutoff frequency of 25 MHz. The photoacoustic image is reconstructed with a standard backprojection algorithm~\cite{xu2005universal}, using a pixel size of $25\ \mu m$ to match the sampling rate of the ultrasound scanner.

The photoacoustic transmission matrix is measured using the protocol described in~\cite{chaigne2014controlling}.
The influence of each of the controllable input modes (SLM pixels) on each of the photoacoustic image pixel is measured by varying the  phase of each SLM pattern (in the Hadamard basis) from 0 to $2\pi$ in 8 steps. 
An unshaped part of the circular input beam serves as a fixed reference field~\cite{popoff2010measuring,chaigne2014controlling}. To increase the measurements SNR, the SLM input modes are in practice measured over a Hadamard basis set~\cite{popoff2010measuring}. A photoacoustic image is acquired for each input SLM phase pattern (as opposed to acquiring only a single one-dimensional time trace as in~\cite{chaigne2014controlling}).
When varying each SLM input pattern phase from 0 to $2\pi$, the intensities values of the photoacoustic images vary according to the cosine of the applied phase [11]. The elements of the photoacoustic transmission matrix are obtained directly by analyzing these cosine modulation phases and amplitudes [11]. To improve the SNR, we analyze the modulation of the maximum intensity values within areas of $125\times 125\ \mu m^2$ that match the acoustic resolution cell size, rather on single pixels, which are considerably smaller.\\

Once the photoacoustic transmission matrix is measured, it can be analyzed to obtained information about the scattering medium and the imaged object. Fig. 2 illustrates how the photoacoustic transmission-matrix approach can improve photoacoustic imaging. Figs 2.b and 2.c illustrate the images that would be obtained with the same imaging system using a uniform illumination. Here, these images were computed by averaging all the images acquired during the matrix measurement process. 
Fig. 2.c is the envelope of Fig 2.b, obtained by computing the modulus of the analytic signal representation of each column (z-axis) of image 2.b. From the photograph shown on Fig. 2.a, it is clear that Figs. 2.b and 2.c only reveals the horizontal (x-axis) structures of the sample. This feature is a standard artefact in photoacoustic imaging, due to the finite aperture of the transducer array~\cite{xu2004reconstructions}: elongated absorbers emits acoustic energy mostly perpendicularly to their main orientation, and therefore only photoacoustic waves emitted from horizontal elements reach the ultrasound array. 
Because the transmission-matrix approach involves measuring a collection of photoacoustic images under several speckle illuminations, its analysis allows to reveal the hidden structures: by analyzing the modulation of the pixels values over the different photoacoustic images, it is possible to reconstruct hidden segments of the leaf skeleton, as shown in Fig.\ref{leaf}.d. 
More specifically, the modulus of the transmission-matrix elements was averaged over the input modes: $I_j= <|T_{i,j}|>_{i}$, where I is the image displayed on Fig.\ref{leaf}.d, and T the photoacoustic transmission-matrix. The index j corresponds to a given pixel of the image and the index i corresponds to the input modes. We thus obtain a modulation image revealing all segments of the leaf skeleton independently of their orientation.
This approach is very similar to that introduced by Gateau et al.\cite{gateau2013improving}, where uncorrelated random speckle patterns were produced using a varying diffuser: images revealing invisible structures were obtained from the measure of the variance between the photoacoustic images obtained under the various random illuminations. This is made possible because the speckle illumination breaks up the elongated parts of the absorbing structure into small isotropic and high frequency sources. 
We emphasize that the image of Fig. 1.d is obtained without any additional measurement, all the information being contained in the transmission-matrix. \\

\begin{figure}[h!]
\centerline{\includegraphics[width=1\columnwidth]{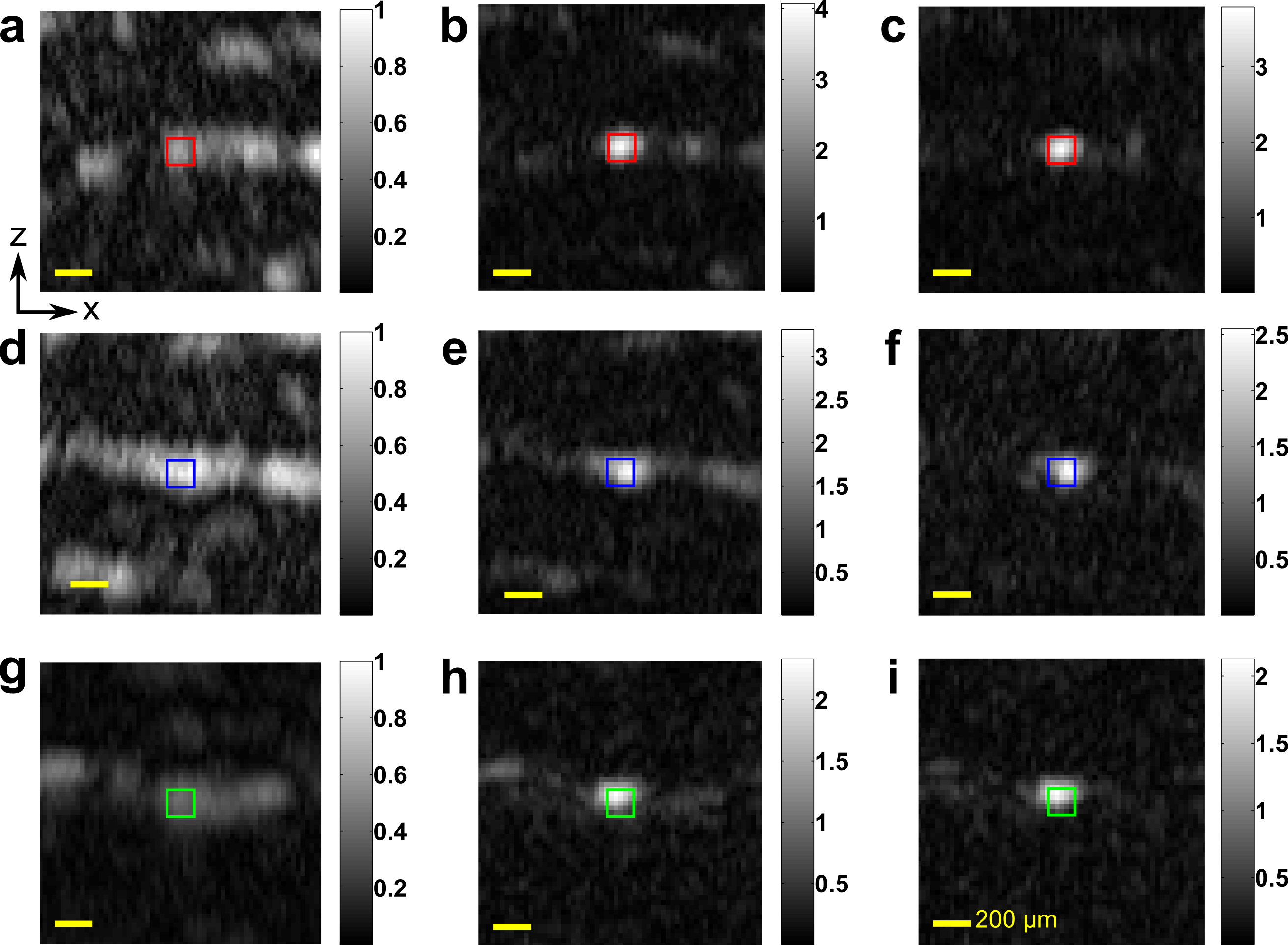}}
\caption{Focusing on \textit{visible} regions: (a) enveloppe of the mean photoacoustic image (zoom of Fig. 2.c). (b) enveloppe of the photoacoustic image when displaying specific focusing phase pattern on the SLM. (c) Enveloppe of the difference between the photoacoustic images corresponding to (b) and (a) (the difference is computed before the enveloppe). (d,e,f) and (g,h,i) are analog to (a,b,c) for different target positions. The colored squares indicate the areas over which the modulation of the photoacoustic signal was measured. The colors correspond to the positions marked on Fig.\ref{leaf}.b.}
\label{focus}
\end{figure}

\begin{figure}[h!]
\centerline{\includegraphics[width=1\columnwidth]{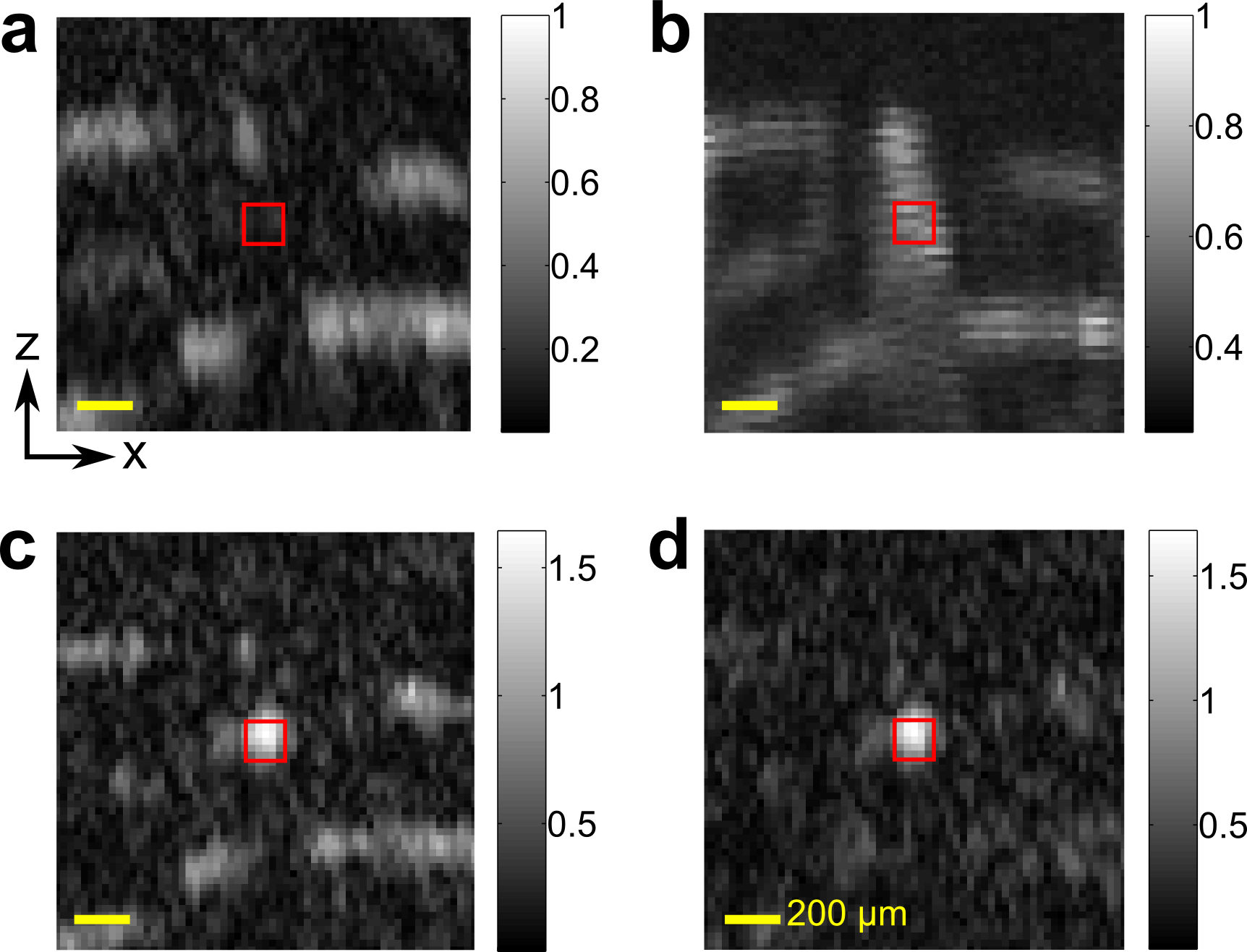}}
\caption{Focusing on \textit{invisible} regions: (a) enveloppe of the mean photoacoustic image (zoom of Fig. 2.c). (b) average of transmission-matrix modulus over the input modes (zoom of Fig 2.d). (c) enveloppe of the photoacoustic image when displaying specific focusing phase pattern on the SLM. (d) Enveloppe of the difference between the photoacoustic images corresponding to (c) and (b) (the difference is computed before the enveloppe).}
\label{invpart}
\end{figure}

As demonstrated in~\cite{chaigne2014controlling}, the photoacoustic transmission-matrix can be used to selectively focus light through a turbid sample. Here we investigate the light-focusing capability of the two-dimensional photoacoustic transmission-matrix for an arbitrary position on the absorbing structure. To this end, for each of the selected desired focusing positions,  we use the transmission-matrix row corresponding to this output mode. We apply the corresponding conjugated phase pattern on the SLM, so that all input modes contribute constructively to maximize the photoacoustic signal  at the target location~\cite{chaigne2014controlling}. The results of focusing on three arbitrary targets on the leaf skeleton fibers are shown in Fig.\ref{focus}. Quantitatively, the signals of the reconstructed photoacoustic image were locally increased by a factor of 6.2 (red target, Fig. 3.b), 3.4 (blue target, Fig. 3.e), and 5.2 (green target, Fig. 3.h). This photoacoustic enhancement factor is defined as the ratio between the maxima over the targeted area (colored square, $125\times 125 \mu m^2 $) of the enhanced (when applying focusing pattern on the SLM) and average (Fig. 2.c) photoacoustic images.
These ratios are within the order of magnitude of the expected enhancement of the optical intensity, defined roughly as $0.5 \times \dfrac{N_{SLM}}{N_{speckles}}$, where $N_{SLM}$ is the number of SLM pixels and $N_{speckles}$ the number of optical speckles grains contained in the targeted acoustic cell \cite{popoff2010measuring}, which optical intensities are simultaneously enhanced. Here, the distance between the scattering sample and the absorbing structure yielded $25 \mu m$ diameter speckle grains on the absorbing structure ($\dfrac{1}{\sqrt{2}} \times$ FHWM of speckle autocorrelation).
The size of the targeted acoustic resolution cells is $125\times 125\mu m^2$. $N_{speckle}$ is thus assessed to be aroud 25, leading to an expected optical enhancement factor around 3. We compute $N_{speckles}$  as a ratio of areas, because the speckle are elongated in the light propagation axis. Moreover, the elevation resolution of the transducer-array is not taken into account here because the absorbing sample is fully contained in the elevation focal zone. In the focusing results presented in Fig.\ref{focus}, we have selected noticeably bright points on the standard photoacoustic image (fig. \ref{leaf}.c) as target points, meaning that focusing was obtained only on rather horizontal veins of the leaf skeleton.  However, using the photoacoustic transmission matrix, light focusing and control can be exerted on normally "invisible" vertical structures as well. In Fig.\ref{invpart}, we demonstrate such focusing on one such portion that is invisible on Fig.\ref{leaf}.c. Even though the pixel values are close to the noise floor around the target point in the standard photoacoustic image (Fig.\ref{invpart}.b), we are able to focus at this location using the same phase-conjugation focusing procedure. The corresponding result is shown on Fig.\ref{invpart}.c. The identification of the appropriate target positions is made possible thanks to the visibility enhancement allowed by the photoacoustic transmission-matrix (Fig. 2.d).
\newline

In summary, we have demonstrated that the implementation of the transmission-matrix approach on a two-dimensional photoacoustic imaging system enables to control light focusing over a large field-of-view without any mechanical scanning process, by parallezing the acquisition of the output modes. We experimentally demonstrate the selective focusing at arbitrary target points on a complex absorbing structure located behind a scattering sample. Moreover, we have shown that structures that are invisible in a standard photoacoustic image due to the limited-view configuration could be revealed by imaging the signals modulation when varying the input modes. In addition, we have demonstrated the capability to focus on these hidden segments using this modulation image to select the target points. We note that the photoacoustic transmission-matrix method suits well the study of complex absorbing samples, as it enables a clear identification of the targets regardless of their orientation relatively to limited-view acquisition and allows to locally enhance photoacoustic signals.\\

Optical focusing was performed up to the acoustic resolution. Further studies are required to understand the effect of wavefront shaping below this current resolution limit. For practical application in deep-tissue imaging, two different challenges emerge. The first is related to the fast decorrelation of optical speckle patterns inside tissue.
The use of higher repetition rate laser and higher damage threshold SLM would improve the speed of the transmission-matrix measurement, which is currently limited to several tens of minutes. The second challenge is the ability to measure the small modulation of the photoacoustic signal from an acoustic-resolution cell containing a large number of optical speckles. One has thus to ensure that the number of optical speckles contained in an acoustic-resolution cell is as low as possible, by using either small absorbers, high frequency transducer, non-linear photoacoustic signal generation or an alternative analysis of the signal fluctuations.\\

This work was funded by the European Research Council (grant number 278025), the Fondation Pierre-
Gilles de Gennes pour la Recherche (grant number FPGG031), and the Plan Cancer 2009-2013 (project Gold
Fever). O. K. acknowledges the support of the Marie Curie intra-European fellowship for career development(IEF).

\vspace{1cm}

\begin{thebibliography}{21}

\bibitem{ntziachristos2010going}
Vasilis Ntziachristos.
\newblock Going deeper than microscopy: the optical imaging frontier in
  biology.
\newblock {\em Nature methods}, 7(8):603--614, 2010.

\bibitem{vellekoop2007focusing}
IM~Vellekoop and AP~Mosk.
\newblock Focusing coherent light through opaque strongly scattering media.
\newblock {\em Optics letters}, 32(16):2309--2311, 2007.

\bibitem{popoff2010measuring}
SM~Popoff, G~Lerosey, R~Carminati, M~Fink, AC~Boccara, and S~Gigan.
\newblock Measuring the transmission matrix in optics: an approach to the study
  and control of light propagation in disordered media.
\newblock {\em Physical review letters}, 104(10):100601, 2010.

\bibitem{mosk2012controlling}
Allard~P Mosk, Ad~Lagendijk, Geoffroy Lerosey, and Mathias Fink.
\newblock Controlling waves in space and time for imaging and focusing in
  complex media.
\newblock {\em Nature photonics}, 6(5):283--292, 2012.

\bibitem{vellekoop2008demixing}
IM~Vellekoop, EG~Putten, A~Lagendijk, and AP~Mosk.
\newblock Demixing light paths inside disordered metamaterials.
\newblock {\em Optics express}, 16(1):67--80, 2008.

\bibitem{hsieh2010imaging}
Chia-Lung Hsieh, Ye~Pu, Rachel Grange, Gr{\'e}goire Laporte, and Demetri
  Psaltis.
\newblock Imaging through turbid layers by scanning the phase conjugated second
  harmonic radiation from a nanoparticle.
\newblock {\em Optics Express}, 18(20):20723--20731, 2010.

\bibitem{xu2011time}
Xiao Xu, Honglin Liu, and Lihong~V Wang.
\newblock Time-reversed ultrasonically encoded optical focusing into scattering
  media.
\newblock {\em Nature photonics}, 5(3):154--157, 2011.

\bibitem{si2012fluorescence}
Ke~Si, Reto Fiolka, and Meng Cui.
\newblock Fluorescence imaging beyond the ballistic regime by
  ultrasound-pulse-guided digital phase conjugation.
\newblock {\em Nature photonics}, 2012.

\bibitem{judkewitz2013speckle}
Benjamin Judkewitz, Ying~Min Wang, Roarke Horstmeyer, Alexandre Mathy, and
  Changhuei Yang.
\newblock Speckle-scale focusing in the diffusive regime with time reversal of
  variance-encoded light (trove).
\newblock {\em Nature Photonics}, 2013.

\bibitem{kong2011photoacoustic}
Fanting Kong, Ronald~H Silverman, Liping Liu, Parag~V Chitnis, Kotik~K Lee, and
  Ying-Chih Chen.
\newblock Photoacoustic-guided convergence of light through optically diffusive
  media.
\newblock {\em Optics letters}, 36(11):2053--2055, 2011.

\bibitem{beard2011biomedical}
Paul Beard.
\newblock Biomedical photoacoustic imaging.
\newblock {\em Interface focus}, 1(4):602--631, 2011.

\bibitem{chaigne2014controlling}
T.~Chaigne, O.~Katz, A.C. Boccara, M.~Fink, E.~Bossy, and S.~Gigan.
\newblock Controlling light in scattering media non-invasively using the
  photoacoustic transmission matrix.
\newblock {\em Nat Photon}, 8(1):58--64, January 2014.

\bibitem{chaigne2013improving}
Thomas Chaigne, Ori Katz, J{\'e}r{\^o}me Gateau, Claude Boccara, Sylvain Gigan,
  and Emmanuel Bossy.
\newblock Improving photoacoustic-guided focusing in scattering media by
  spectrally filtered detection.
\newblock {\em arXiv preprint arXiv:1310.7535}, 2013.

\bibitem{2013arXiv1310.5736C}
D.~B. {Conkey}, A.~M. {Caravaca-Aguirre}, J.~D. {Dove}, H.~{Ju}, T.~W.
  {Murray}, and R.~{Piestun}.
\newblock {Super-resolution photoacoustic imaging through a scattering wall}.
\newblock {\em ArXiv e-prints}, October 2013.

\bibitem{caravaca2013high}
Antonio~M Caravaca-Aguirre, Donald~B Conkey, Jacob~D Dove, Hengyi Ju, Todd~W
  Murray, and Rafael Piestun.
\newblock High contrast three-dimensional photoacoustic imaging through
  scattering media by localized optical fluence enhancement.
\newblock {\em Optics express}, 21(22):26671--26676, 2013.

\bibitem{gateau2013improving}
J{\'e}r{\^o}me Gateau, Thomas Chaigne, Ori Katz, Sylvain Gigan, and Emmanuel
  Bossy.
\newblock Improving visibility in photoacoustic imaging using dynamic speckle
  illumination.
\newblock {\em Optics Letters}, 38(23):5188--5191, 2013.

\bibitem{vellekoop2010exploiting}
IM~Vellekoop, Ad~Lagendijk, and AP~Mosk.
\newblock Exploiting disorder for perfect focusing.
\newblock {\em Nature Photonics}, 4(5):320--322, 2010.

\bibitem{jose2013speed}
Jithin Jose, Rene G.~H. Willemink, Wiendelt Steenbergen, C.~H. Slump, Ton~G.
  van Leeuwen, and Srirang Manohar.
\newblock Speed-of-sound compensated photoacoustic tomography for accurate
  imaging.
\newblock {\em Medical Physics}, 39(12):7262--7271, 2012.

\bibitem{huang2013improving}
Bin Huang, Jun Xia, Konstantin Maslov, and Lihong~V Wang.
\newblock Improving limited-view photoacoustic tomography with an acoustic
  reflector.
\newblock {\em Journal of biomedical optics}, 18(11):110505--110505, 2013.

\bibitem{xu2005universal}
Minghua Xu and Lihong~V Wang.
\newblock Universal back-projection algorithm for photoacoustic computed
  tomography.
\newblock {\em Physical Review E}, 71(1):016706, 2005.

\bibitem{xu2004reconstructions}
Yuan Xu, Lihong~V Wang, Gaik Ambartsoumian, and Peter Kuchment.
\newblock Reconstructions in limited-view thermoacoustic tomography.
\newblock {\em Medical Physics}, 31:724, 2004.

\end{thebibliography}

\end{document}